\documentclass[twocolumn,prb,showpacs]{revtex4}
\usepackage{graphicx}
\usepackage{epsfig}
\usepackage{color}

\makeatother
\begin{document}

\title{Zigzag spin chains with antiferromagnetic-ferromagnetic interactions: Transfer-matrix renormalization group study}
\author{H. T. Lu,$^{1,2}$ Y. J. Wang,$^{3}$ Shaojin Qin,$^{2}$ T. Xiang$^{2}$}

\affiliation{${}^{1}$School of Physics, Peking University, Beijing 100871, China\\
${}^{2}$Institute of Theoretical Physics and Interdisciplinary Center
of Theoretical Studies, Chinese Academy of Sciences, Beijing 100080, China \\
${}^{3}$Department of Physics, Beijing Normal University, Beijing 100875, China}

\date{}

\begin{abstract} 
Properties of the zigzag spin chains with various nearest-neighbor and
next-nearest-neighbor interactions are studied by making use of the
transfer-matrix renormalization group method. Thermodynamic quantities of the
systems (temperature dependence of the susceptibility and the specific heat),
as well as the field dependence of the magnetization are analyzed numerically
with a high accuracy in the thermodynamic limit. The results have been
compared with the recent experimental data on Rb$_2$Cu$_{2}$Mo$_{3}$O$_{12}$.
\end{abstract}


\maketitle

\section{Introduction}
\label{sec:intro}

Spin models with charge degrees of freedom frozen, as an effective low-energy
description for insulating systems, have been proved to be a fruitful resource
of fundamental concepts and principles in condensed matter physics and other
related fields. Besides theoretical explorations, advanced probe tools and
measurements, combined with multifarious natural and artificial materials,
such as ceramic or organic compounds, and optical lattices provide remarkable
practical platforms for theoretical investigations. The simplest
one-dimensional (1D) quantum spin model is the nearest-neighbor (NN) Heisenberg
model:
\begin{equation}
H=J\sum_i\mathbf{S}_{i}\cdot\mathbf{S}_{i+1} \,,
\label{eq:1}
\end{equation}
where $\mathbf{S}_{i}$ is the quantum spin operator. From Bethe's seminal
paper\cite{Bethe} and successive works, we know that the ground state of a
spin-$\frac{1}{2}$ antiferromagnetic (AF) $\left(J>0\right)$ chain is a
singlet $\left(\mathbf{S}_{\text {tot}}=0\right)$ and has quasi-long-range
order with algebraically decaying spin correlations. The gapless spectrum
contains no single-particle excitations and is instead a continuum of states.
The elementary excitations are called spinons, which carry spin $1/2$ and
appear only in pairs in all physical states with integer total spin. This
picture is qualitatively different from the prediction of a spin-wave theory,
which is usually effective in higher-dimensional systems. The gapless behavior
is special for half-integer Heisenberg spin chains, while for integer ones,
Haldane conjectured that there exists a finite gap.\cite{Haldane2}
Experimental, numerical, and theoretical studies have confirmed this conjecture
for $S=1$ and some other higher spin values.\cite{H_conjecture} On the other
hand, half-integer spin chains can be driven to a gapped phase by frustrations.
Throughout this paper, we confine our discussions to the spin-$1/2$
case.

\begin{figure}
\centering
\includegraphics[width=8cm]{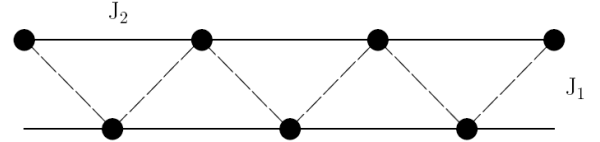}
\caption{\label{zigzag}Heisenberg zigzag spin chains.}
\end{figure}

A straightforward generalization of the NN Heisenberg model is to
include the next-nearest-neighbor (NNN) interactions:
\begin{equation}
H=\sum_{i=1}^{N}\left(J_{1}\mathbf{S}_{i}\cdot\mathbf{S}_{i+1}
+J_{2}\mathbf{S}_{i}\cdot\mathbf{S}_{i+2}\right) \,, \label{eq:2}
\end{equation}
where $J_{1}$ or $J_{2}$ is NN or NNN exchange interaction constant, and
periodic boundary conditions are implied. This is usually called the
$J_{1}$-$J_{2}$ model, which can be also considered as a model for a zigzag
spin chain, as shown in Fig.~\ref{zigzag}.

The $J_{1}$-$J_{2}$ model has been investigated theoretically over
the decades. With $J_{2}>0$, it is a frustrated (competing) system,
irrespective of the sign of $J_{1}$. When $J_{1}>0$ and $J_{2}>0$
(AF-AF), the ground state is a spin liquid. The increase of the
ratio of the coupling constants $\alpha (\equiv J_{2}/J_{1})$
induces an infinite-order phase transition from a gapless state to a
gapful dimerized state.\cite{Haldane,WhiteAffleck} The critical
point $\alpha_{c}$ is numerically estimated to be
$\approx0.241$.\cite{Eggert} When $\alpha$ is further increased to
the so-called Majumdar-Ghosh (MG) point at 1/2, the ground state is
the products of singlet pairs formed by nearest neighboring
spins.\cite{Majumdar,ShastrySutherland} It is two-fold degenerate
since the $Z_{2}$ symmetry of translations by one site is
spontaneously broken. When $J_{1}<0$ and $J_{2}>0$ (F-AF) with
$-1/4<\alpha\leq0$, the ground state is fully ferromagnetic (FM),
and becomes an $(S=0)$ incommensurate state for
$\alpha<-1/4$.\cite{Tonegawa} It is suggested that in this
incommensurate state the gap is strongly suppressed.\cite{Qin} For
$\alpha=-1/4$, the exact ground state can be shown\cite{Hamada} to
have a $(N+2)$-fold degeneracy, comprised by $S=0$ and $S=N/2$
states (with $N$ the lattice size). When $J_{1}>0$ and $J_{2}<0$
(AF-F), the system is believed to be in a gapless antiferromagnetic phase
for any permissible values of $J_{1}$ and $J_{2}$.

There are many papers contributing to the AF-AF case and the related extended
models. The other two cases (F-AF and AF-F), although having caught relatively
less attention, also show very interesting phenomena, which we will mainly
deal with in this paper. Aside from the general aspect of theoretical
interest, especially for understanding the roles played by frustration and
incommensurability, an additional motivation to study this system lies in the
fact that physically, it is believed that a large class of copper oxides can be
essentially described by the $J_{1}$-$J_{2}$ model.

\begin{figure}[b]
\centering
\includegraphics[%
  bb=0bp 0bp 320bp 200bp,
  width=3.1cm]{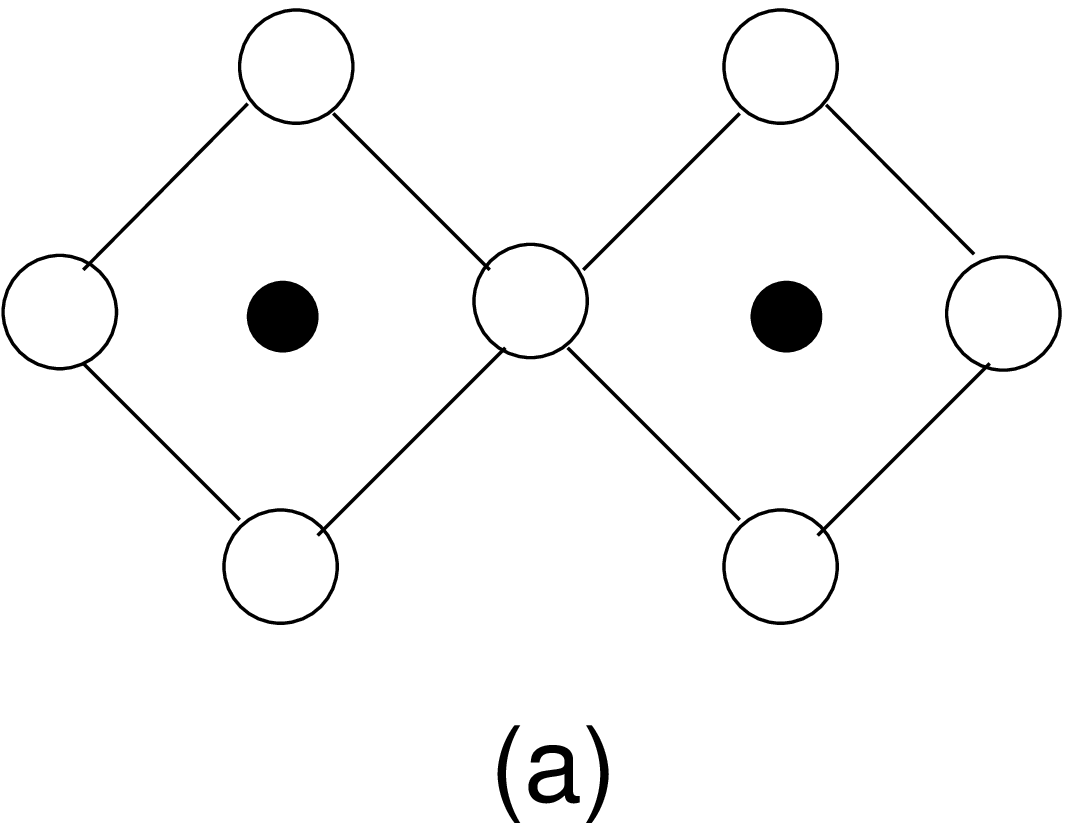}
  \hspace{0.1cm}
  \includegraphics[%
  bb=0bp 0bp 330bp 248bp,
  width=3.2cm]{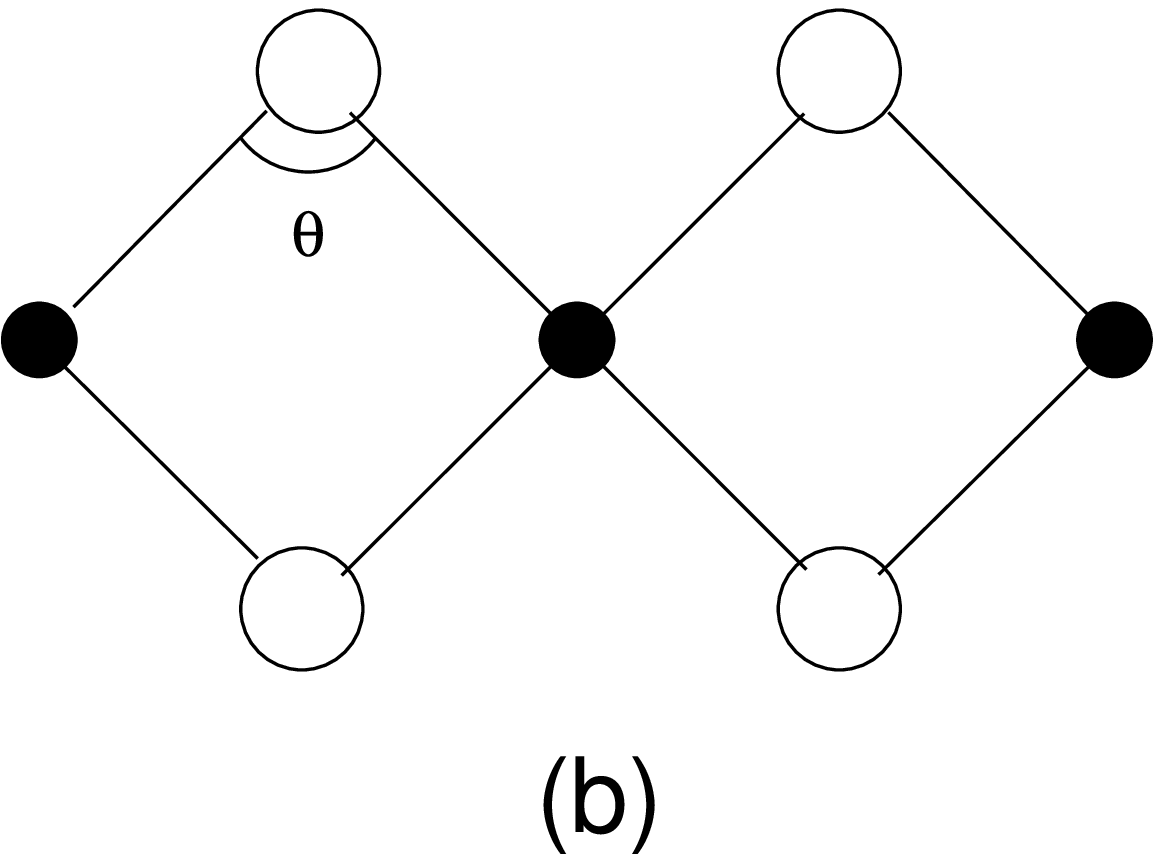}
  \includegraphics[%
  bb=0bp 0bp 100bp 70bp,
  width=1.5cm]{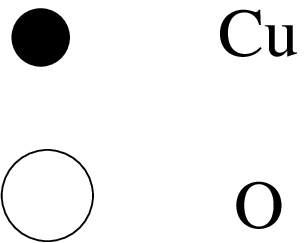}

\includegraphics[%
  bb=0bp 0bp 232bp 310bp,
  width=2.6cm]{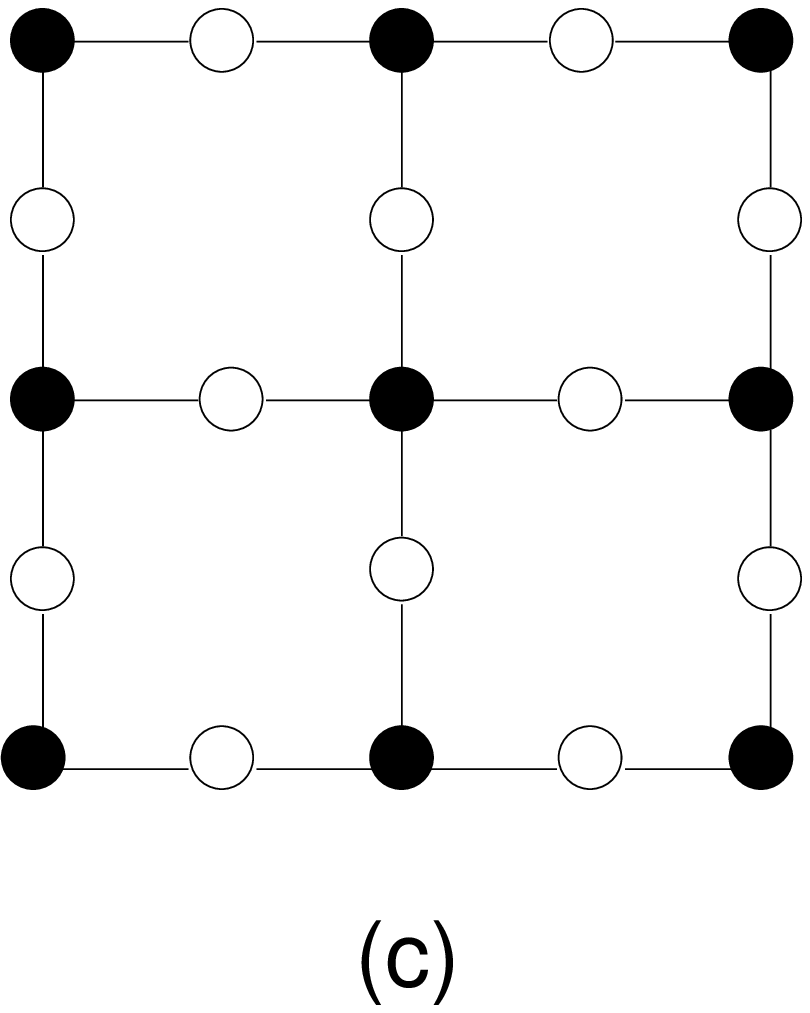}
  \hspace{0.1cm}
  \includegraphics[%
  bb=0bp 0bp 390bp 290bp,
  clip,
  width=4.7cm]{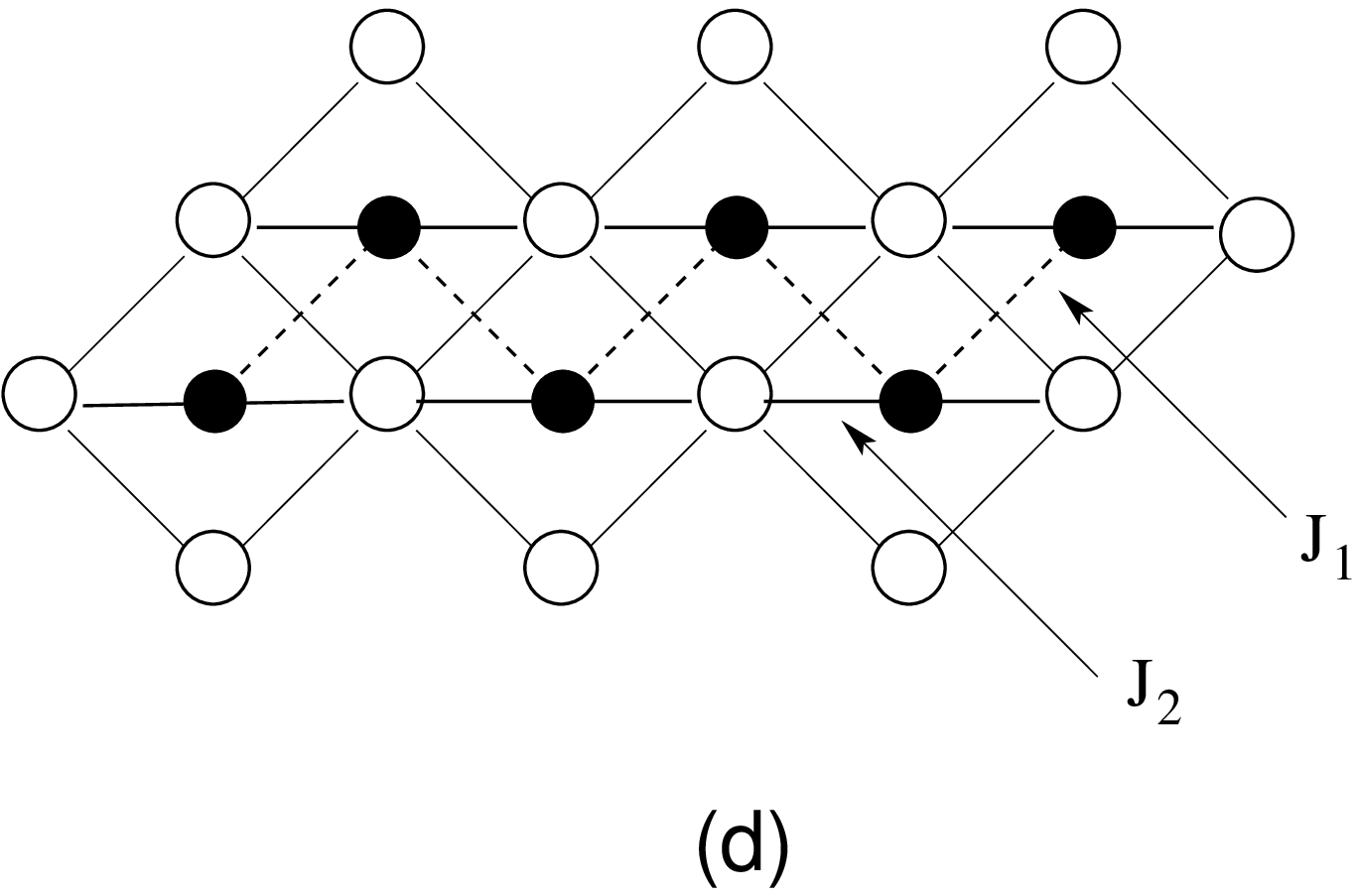}
\caption{\label{EC}Various ways of linking CuO$_{4}$ plaquettes. (a)
The corner-sharing chain, as in Sr$_2$CuO$_3$; (b) the edge-sharing
chain, as in CuGeO$_3$, Li$_2$CuO$_2$, and Rb$_2$Cu$_2$Mo$_3$O$_{12}$;
(c) CuO$_{2}$ plane expanded by corner-sharing chains, as in
cuprates; (d) the zigzag chain in SrCuO$_{2}$.}
\end{figure}


Copper oxides are excellent model systems for low dimensional
spin-$\frac{1}{2}$ quantum magnets, where magnetic Cu$^{2+}$ ions carry $1/2$
spins. The basic building blocks are CuO$_{4}$ plaquettes and there are three
ways of linking these fundamental units. One is adjacent squares sharing their
corners, as shown in (a) of Fig.~\ref{EC}. A typical example is Sr$_2$CuO$_3$.
These corner-sharing chains can expand in the plane to form a CuO$_{2}$ sheet,
constituting the basic structure in cuprates. The dominant interaction in a
corner-sharing chain is the NN superexchange.  Linear Cu-O-Cu bonds along the
spin chains give rise to a large antiferromagnetic NN exchange coupling. As in
Sr$_{2}$CuO$_{3}$, the NN coupling constant is estimated to be $2100\pm200$
K.\cite{SrCuO_2} The second kind is edge-sharing, as shown in Fig.~\ref{EC}
(b). Because of the nearly 90$^{\circ}$ Cu-O-Cu bond in the edge-sharing
squares, an O$2p_{\sigma}$ orbital hybridizing with a $3d$ orbital of Cu ion
is almost orthogonal to that of the next Cu ion. The NN interaction $J_{1}$
can vary from antiferromagnetic to ferromagnetic, as the angle $\theta$ of the
Cu-O-Cu bond approaches 90$^{\circ}$ from a larger value.\cite{Mizuno} This
nearly orthogonality makes the NN coupling in the edge-sharing case more than
an order of magnitude smaller than the corner-sharing. The sign and the
absolute value of the NN interaction depend sensitively on the bond angle
$\theta$ and the distance between copper and oxygen ions. We refer readers to
Table I of Ref.~\onlinecite{Hase} for details. The third configuration is the
combination of corner-sharing and edge-sharing in one spin chain
simultaneously, such as in SrCuO$_{2}$, as shown in Fig.~\ref{EC} (d). In
SrCuO$_{2}$, the NNN coupling, which is the superexchange interaction through
the linear Cu-O-Cu, is almost ten times greater than the NN one through
sharing the edges.\cite{SrCuO2}

In corner-sharing chains, the NNN interaction can often be neglected safely
due to its small magnitude relative to the NN interaction.  For the
edge-sharing case, the situation is quite different. The NNN interaction
$J_{2}$ through the Cu-O-O-Cu path is generally antiferromagnetic, usually
with a magnitude of a few tens Kelvin.  Despite its smallness, $J_{2}$ has a
pronounced effect on the physical properties of these systems since the NN
coupling is also small. Therefore, edge-sharing copper oxides provide abundant
experimental materials for studying the zigzag spin chain model, which can
cover a large region of the parameter space.

On the other hand, at low temperatures, besides intrachain couplings $J_{1}$
and $J_{2}$, other forms of interactions often become relevant, driving a
system to various phases with the decrease of temperature. For example,
spin-phonon interactions can induce a spin-Peierls instability, as in
CuGeO$_{3}$.\cite{CuGeO3} If interchain interactions are strong enough, an
antiferromagnetically long-ranged order will appear below the N\'eel
temperature $T_{N}$, as in Ca$_{2}$Y$_{2}$Cu$_{5}$O$_{10}$,\cite{CaYCuO}
La$_{6}$Ca$_{8}$Cu$_{24}$O$_{41}$,\cite{LaCaCuO} and
Li$_{2}$CuO$_{2}$.\cite{Li2CuO2_4} Another interesting case is
LiCu$_{2}$O$_{2}$,\cite{LiCu2O2} which undergoes a transition to a magnetic
helix state at low temperatures. By a comparative study on the similar oxide
NaCu$_{2}$O$_{2}$, it has been shown that interchain interactions over a few
chains should be incorporated to explain the experimental results.\cite{NaCuO}
These facts reveal the complexity of the interactions underlying edge-sharing
copper oxides at low temperatures.

Since the effect of NNN interaction is important for edge-sharing copper oxide
chains, the study on the $J_{1}$-$J_{2}$ model not only has its own
theoretical meaning, but also obtains a connection with practical materials. In
this paper, in the thermodynamic limit and extending to the low-temperature
region, we use the transfer-matrix renormalization group (TMRG)
method~\cite{Bursill2,XiangWang2,XiangWang} to study the thermodynamic
properties of the $J_{1}$-$J_{2}$ model with antiferromagnetic-ferromagnetic
interactions.

The TMRG is a powerful numerical tool for studying the thermodynamic
properties of 1D quantum systems.  It starts by expressing the
partition function as a trace on the product of the transfer matrix
$\mathcal{T}_{M}$ using the Trotter-Suzuki decomposition.
\begin{equation}
Z=\mathrm{Tr}e^{-\beta
H}=\lim_{M\rightarrow\infty}\mathrm{Tr}\mathcal{T}_{M}^{N/2},
\end{equation}
where $M$ is the Trotter number, and $N$ is the total cell number in the
lattice, $\tau=\beta/M$. The definition of $\mathcal{T}_{M}$ can be found from
Refs. \onlinecite{Bursill2, XiangWang2, XiangWang}.  For the $J_1$-$J_2$ model
considered here, each cell consists of two adjoining spins.  In the
thermodynamic limit $N\rightarrow\infty$, the free energy $f$, internal energy
$u$, and uniform magnetization $m_z$ can be expressed by the maximum
eigenvalue $\lambda_{\mbox{max}}$ and the corresponding left
$\langle\psi^{L}\mid$ and right $\mid\psi^{R}\rangle$ eigenvectors of the
transfer matrix $\mathcal{T}_{M}$:
\begin{eqnarray}
f & = &-\lim_{N\rightarrow\infty}\frac{1}{N\beta}\ln Z
= -\frac{1}{2\beta}\lim_{M\rightarrow\infty}\ln\lambda_{\mbox{max}},\\
u & = &\frac {\langle\psi^{L}\mid\tilde{\mathcal{T}}_{U}
\mid\psi^{R}\rangle}{\lambda_{\mbox{max}}},\\
m_z & = & \frac{\langle\psi^{L}\mid\tilde{\mathcal{T}}_{M}
\mid\psi^{R}\rangle}{\lambda_{\mbox{max}}},
\end{eqnarray}
where the definition of the transfer matrices $\tilde{\mathcal{T}}_{U}$ and
$\tilde{\mathcal{T}}_{M}$, which are similar to $\mathcal{T}_{M}$, can also be
found in Refs. \onlinecite{Bursill2, XiangWang2, XiangWang}.  The specific
heat and magnetic susceptibility can then be calculated by numerical
derivatives of $u$ and $m_z$, respectively,
\begin{eqnarray}
C & =&\frac{\partial u}{\partial T},\\
\chi & =&\frac{\partial m_z}{\partial H}.
\end{eqnarray}
In our numerical simulation, $\tau =0.05$, the error caused by the
Trotter-Suzuki decomposition is less than $10^{-3}$. During the TMRG
iterations, $60-80$ states are retained and the truncation error is less
than $10^{-4}$ down to $k_BT\sim0.01J$.

The outline of this paper is as follows: Section \ref{sec:FAF} and
Sec.~\ref{sec:AFF} are devoted to discussions on F-AF and AF-F cases,
respectively. Section \ref{sec:EXP} shows our numerical results compared with
the newly experiments on Rb$_{2}$Cu$_{2}$Mo$_{3}$O$_{12}$.~\cite{Hase} We
conclude with a brief summary in Sec.~\ref{sec:SUM}.

\section{the F-AF case}
\label{sec:FAF}

As mentioned in the previous section, at the critical point
$\alpha_{c}=-1/4$, two distinct configurations with the energy
$E_{g}=-3N|J_{1}|/16$ are the ground states, where $N$ is the
lattice size.\cite{Hamada} One is fully ferromagnetic with
$S_{\text{tot}}=N/2$, the other is a singlet state with
$S_{\text{tot}}=0$. The state vector for the latter can be expressed
as
\[
\Phi=\sum\left[i,j\right]\left[k,l\right]\left[m,n\right]\cdots \,,
\]
 where the summation is made for any combination of spin sites under
the condition that $i<j$, $k<l$, $m<n$, $\cdots$, and
$\left[i,j\right]$ denotes the singlet pair. This is also called a
uniformly distributed resonant-valence-bond (UDRVB) state.


In the region $0\ge\alpha>-1/4$, the ground state lies in the subspace
$\mathbf{S}_{\text{tot}}=N/2$ with the degeneracy $N+1$. The ground state
energy $E_{g}=-N|J_{1}|\left(1+\alpha\right)/4$.  For $\alpha<-1/4$, the
ground state lies in the subspace
$\mathbf{S}_{\text{tot}}=S_{\text{tot}}^{z}=0$ and the lattice translational
symmetry is thought to be broken.  For the critical point $\alpha_{c}=-1/4$,
besides the ferromagnetic configuration, the UDRVB state restores the lattice
translational symmetry.

When $\alpha<-1/4$, whether the system is gapped or gapless is an interesting
and controversial issue. When $J_{2}\gg|J_{1}|$, it is appropriate to regard
the model as two antiferromagnetic spin chains coupled by a weak zigzag
interchain interaction $J_{1}$. This coupling can be expressed by the
current-current interaction\cite{WhiteAffleck,Allen2} in terms of the
Wess-Zumino-Witten fields (see, e.g., Ref.~\onlinecite{XLIX}). If $J_{1}>0$,
by renormalization group (RG) analysis, this interaction is marginally
relevant and produces an exponentially small gap
$\Delta\propto\exp\left(-\text{const} J_{2}/J_{1}\right)$, which leads to a
spontaneously dimerized ground state.\cite{WhiteAffleck,Allen2} While at the
ferromagnetic side, i.e. $J_{1}<0$, the model was believed to be gapless
because the current-current interaction renormalizes logarithmically to
zero.\cite{WhiteAffleck} It was conjectured that the ferromagnetic model is
critical with different velocities for the spin-singlet and spin-triplet
excitations.\cite{Allen2} Afterward, Nersesyan {\it et al.}\cite{Nersesyan} found
that in addition to the current-current interaction, a ``twist'' term
associated with the staggered component of the spin operators arises in the
zigzag chains.  Due to this parity-breaking term, the critical point $J_{1}=0$
is unstable both in the ferromagnetic $\left(J_{1}<0\right)$ and
antiferromagnetic $\left(J_{1}>0\right)$ regions.\cite{Nersesyan,Cabra}

The phase diagram for the F-AF model in an external magnetic field was
discussed by Chubukov.~\cite{Chubukov} In addition to the ferromagnetic phase,
two different biaxial and uniaxial spin nematic phases are mapped out. In
these nematiclike phases there is an extra symmetry breaking of reflections
about a bond or about a site. In the absence of external field, with the
decrease of $\alpha$ starting from $-1/4$, the system develops from the chiral
biaxial spin nematic phase to the dimerized uniaxial spin nematic phase at
$\alpha\simeq-0.385$. The Lieb-Schultz-Mattis theorem~\cite{LSM} states that a
half-integer spin chain with essentially any reasonably local Hamiltonian
respecting translational and rotational symmetries either has a zero gap (i.e.
"mass") or else has degenerate ground states, corresponding to a spontaneously
broken parity. If the phase diagram is correct, we could expect an energy gap
at $\alpha<-1/4$.

On the other hand, numerical analysis has shown a complicated size dependence
of the ground-state energy and correlation function in the region
$\alpha<-1/4$.\cite{Tonegawa,Bursill} This phenomenon, combined with the fact
of slow convergence and no detection of energy gap at the resolution of the
numerical simulations indicates an unusually long correlation length in the
F-AF chain.

By taking into account the twist term, the RG analysis\cite{Qin} found that
although the unstable RG flow around the critical point $J_{1}=0$ produces an
energy gap in the ferromagnetic coupling as well as the antiferromagnetic one,
in the ferromagnetic side, due to the existence of a marginally relevant fixed
line, the gap is strongly reduced. In an extended region for the $J_{1}$ of
order one, the correlation length can be extremely large and the gap, if
exists, is so strongly suppressed that numerical methods can not detect it.

Another closely related topic worth mentioning is the incommensurability in
the zigzag spin chains. 
For simple antiferromagnetic Heisenberg spin-$1/2$ chains and ladders,
the well known mechanisms for generating incommensurabilities are via external
magnetic fields or Dzyaloshinskii-Moria interaction. While in the zigzag spin
chain, it has been found that the frustrated interaction $J_{2}>0$ can also
produce incommensurability.\cite{WhiteAffleck,Bursill}

In classical picture, by regarding spin operators $\mathbf{S}_{i}$
as classical vectors, the energy per site can be expressed as
\begin{equation}
E\left(\theta\right)=J_{1}\cos\theta+J_{2}\cos2\theta \label{eq:3}
\end{equation}
for Hamiltonian (\ref{eq:2}). The pitch angle is given by
\begin{equation}
\cos\theta=-J_{1}/4J_{2} \label{eq:4}
\end{equation}
for $|\alpha|=|J_{2}/J_{1}|\ge1/4$. As $J_{1}>4J_{2}\ge0$, $\theta=\pi$;
$-J_{1}>4J_{2}\ge0$, $\theta=0$. At the special point $J_{1}=0$,
$\theta=\pi/2$, and the
deviations from the normal values ($\pi$ for the AF-AF case and $0$ for the
F-AF) begin to occur at $|\alpha|=1/4$. In quantum level, the characteristic momentum
$Q$ of the spin-spin correlation function which maximizes the static structure
factor is either $\pi$ or $0$, corresponding to the antiferromagnetic or
ferromagnetic spin chain with $J_{2}=0$, respectively. With the increase of
$J_{2}$, $Q$ departs from its usual value. Numerical
simulations~\cite{WhiteAffleck,Bursill} display that for the AF-AF case, $Q$
deviates from $\pi$ with the increase of $J_{2}$ after crossing the MG point
$\left(J_{2}=J_{1}/2\right)$ where the dimerization has already taken place.
On the other side, the departure of $Q$ from $0$ sets in at the critical point
$J_{2}=-J_{1}/4$ for the F-AF case. This asymmetry may be understood on
account of the effects of the current-current interaction and the twist term.

It was shown that in the presence of exchange anisotropic, the twist term
induces incommensurabilities in the spin correlations.\cite{Nersesyan} We have
reason to expect this to hold true even in the SU(2) symmetric (i.e.
isotropic) case.\cite{Allen,Aligia} In the AF-AF isotropic case, both the
twist term and the current-current interaction diverge simultaneously to reach
the strong coupling phase as the RG flows to $J_{2}\gg J_{1}$.\cite{Allen} The
pure current-current interaction induces massive spinons which can be regarded
as quantum dimerization kinks, driving the zigzag spin chain into the
dimerized phase. The twist term appearing in the zigzag case is merely to
shift the minimum of the two-spinon continua to incommensurate wave numbers,
which does not alter the massive spinon picture qualitatively.\cite{Allen} If
these two terms become relevant at different points, we may observe the
emergence of dimerization and incommensurability one after the other, as in
the AF-AF case. For the F-AF case, the current-current interaction is
marginally irrelevant since it renormalizes logarithmically to
zero.\cite{WhiteAffleck} The twist term becomes dominative. We can observe the
incommensurability only, and hardly detect the gap following with the
dimerization. But there is not a simple way to separate the effects of the
current-current and twist interactions in the isotropic zigzag spin chains as
in the anisotropic ones.  There is still much theoretical work to be
accomplished.

Since the TMRG gives results for observable quantities in the thermodynamic
limit, we are able to exploit it to study the bulk properties of the system
without worrying the finite size effect.  The complicated size dependence of
the ground-state energy and correlation function found in the region
$\alpha<-1/4$~(Refs. \cite{Tonegawa,Bursill}) may be avoided.

\begin{figure}[t]
\centering
\includegraphics[width=8cm]{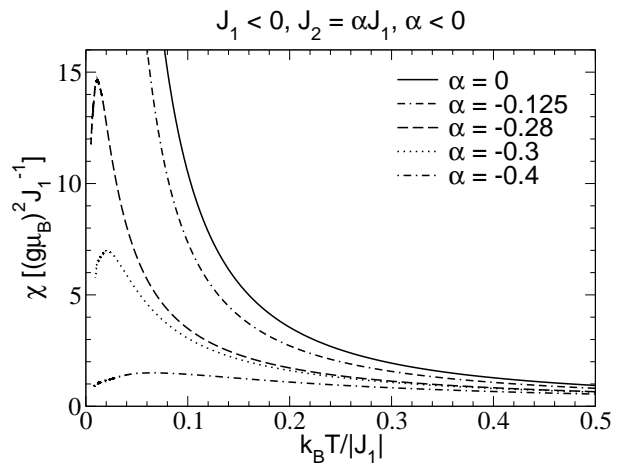}
\caption{\label{chi_FAF}Temperature dependence of the uniform susceptibility
at various $\alpha$ for the F-AF case.}
\vspace{0.2cm}
\end{figure}

\begin{figure}[t]
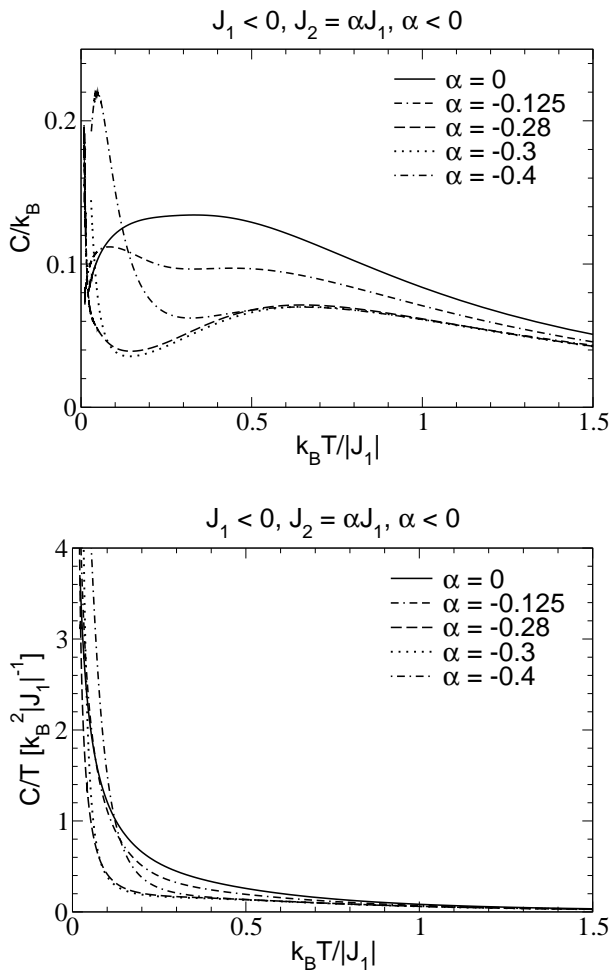

\centering
\includegraphics[width=8cm]{heat_FAF.eps}\\
\vspace{0.5cm}
\includegraphics[width=8cm]{CT_FAF.eps}
\caption{\label{C_FAF}The specific heat $C$ and heat
coefficient $C/T$ at various $\alpha$ for the F-AF case.}
\end{figure}

Figures~\ref{chi_FAF} and \ref{C_FAF} show the TMRG results on the
temperature dependence of the susceptibility $\chi$, specific heat $C$, and
heat coefficient $C/T$ at various $\alpha$ for the F-AF case. At $\alpha=0$,
the model reduces to the case of ferromagnetic spin chain. The curves of
$\alpha=0$ describe the properties of $\chi$ and $C$ for a ferromagnetic spin
chain, which behave in the low temperature limit:
\begin{eqnarray*}
\chi & \sim &T^{-2} \,,\\
C & \sim & \sqrt{T} \,.
\end{eqnarray*}

At $\alpha>-1/4$, the temperature dependence of $\chi$ always diverges,
indicating that the system lies in a ferromagnetic state.  At the critical
point $\alpha=-1/4$, a phase transition from the ferromagnetic state to the
singlet state takes place. In the region $\alpha<-1/4$, remarkable
suppressions of the susceptibility can be observed, and with the decrease of
$\alpha$, the peaks of $\chi$ move to higher temperatures with its heights
decreased rapidly.

For the temperature dependence of the specific heat, the most remarkable
feature is the development of a double-peak structure.  With the decrease of
$\alpha$, at intermediate temperatures, a broad maximum is maintained and its
height lowered. A relatively sharp peak at low temperatures, induced by the
NNN AF interaction $J_2$, appears and develops, and its position moves to
higher temperatures. We expect that this peak would approach the maximum of
$C(T)$ for a pure AF Heisenberg chain.~\cite{Johnston} The drastic change on
the shape of the curves between $\alpha=-0.125$ and $\alpha=-0.28$ implies a
phase transition. The similar double-peak structure for the $J_1$-$J_2$ model
in the F-AF case has also been found in Ref.~\onlinecite{Thanos}. By applying
a method of hierarchy of algebras, the authors calculated thermodynamic
quantities of the linear ring of size 16 described by the same model
(\ref{eq:2}). Our results are qualitatively in agreement with theirs.
Recently, this double-peak structure in the specific heat for the moderate
value $\alpha=-1/3$ has also been confirmed by exact diagonalization
methods.~\cite{Meisner06} Furthermore, for various $\alpha$ we have
considered, $C/T$ as a function of $T$ decreases monotonously down to
$T\sim0.03$. This fact reflects a high density of low-excitation states in
this region, and confirms the results obtained from the density-matrix
renormalization group calculations.\cite{Qin}

From the numerical results above, we find that the existence of gap near the
critical point $\alpha_{c}$ is still a question. The behaviors of
susceptibility and specific heat indicate that if a gap exists near
$\alpha<-1/4$, it should be very small. The properties of the excitations near
and far from $\alpha_{c}$ need to be investigated further.

\section{the AF-F case}
\label{sec:AFF}

In the AF-F case with $J_{1}>0$ and $J_{2}<0$, the behavior of the zigzag spin
chain is believed to be antiferromagnetic with no gap for any permissible
values of $J_{1}$ and $J_{2}$.\cite{Bursill} The arguments are based on a
simple spin-wave analysis. The spectrum of spin-wave excitations is given by
\begin{equation}
\epsilon_{k}=2S\sqrt{\lambda_{k}^{2}-\gamma_{k}^{2}} \,,
\label{eq:5}
\end{equation}
where $S=1/2$, $\alpha=J_{2}/J_{1}<0$,
$\lambda_{k}=-2\alpha\sin^{2}k+1$, $\gamma_{k}=\cos k$.
$\epsilon_{k}$ is linear as $k\rightarrow0$:
\begin{equation}
\epsilon_{k}\sim\upsilon k
\end{equation}
with $\upsilon=2S\sqrt{-4\alpha+1}$. Although the spin-wave theory is not
quite correct for antiferromagnetic spin chains, it is still insightful in
shedding light on the qualitative picture of the excitations around the
characteristic momentum $k=0,\pi$. We see that for the AF-F case, there also
exist gapless excitations at $k=0,\pi$.

According to Lieb's discussion,\cite{LiebMattis} Hamiltonian (\ref{eq:2}) with
$J_{1}>0, J_{2}<0$ describes spin systems at a bipartite lattice. The absolute
ground state of the AF-F spin chain should lie in the $\mathbf{S}=0$ sector.
On the other hand, the relative ground state in each subspace $V(M)$ of the
total magnetic quantum number $M$ is unique. Thus we deduce that the absolute
ground state in the AF-F case is nondegenerate. The Lieb-Schultz-Mattis
theorem asserts that for a half-integer spin chain with reasonably local
Hamiltonian respecting translational and rotational symmetries, either the
parity is spontaneously broken in the ground state or else there are gapless
excitations of odd parity, under the condition that the ground state is
rotational invariant.\cite{LSM} Combining these two statements, we expect the
system should be gapless.

It is not trivial to make the above statement completely rigorous, considering
the relative ground state in the subspaces $V(M)$ may not be unique in the
thermodynamic limit $N\rightarrow\infty$. In order to confirm the conclusion,
we calculate the temperature dependence of the susceptibility and specific
heat for various $\alpha$ by using the TMRG method.

\begin{figure}[t]
\centering
\includegraphics[width=8cm]{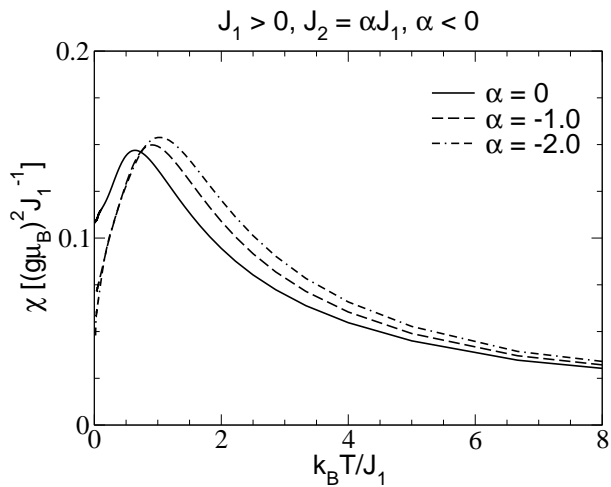}
\caption{\label{chi_AFF}Temperature dependence of the uniform susceptibility
at various $\alpha$ for the AF-F case.}
\vspace{0.2cm}
\end{figure}

\begin{figure}[t]
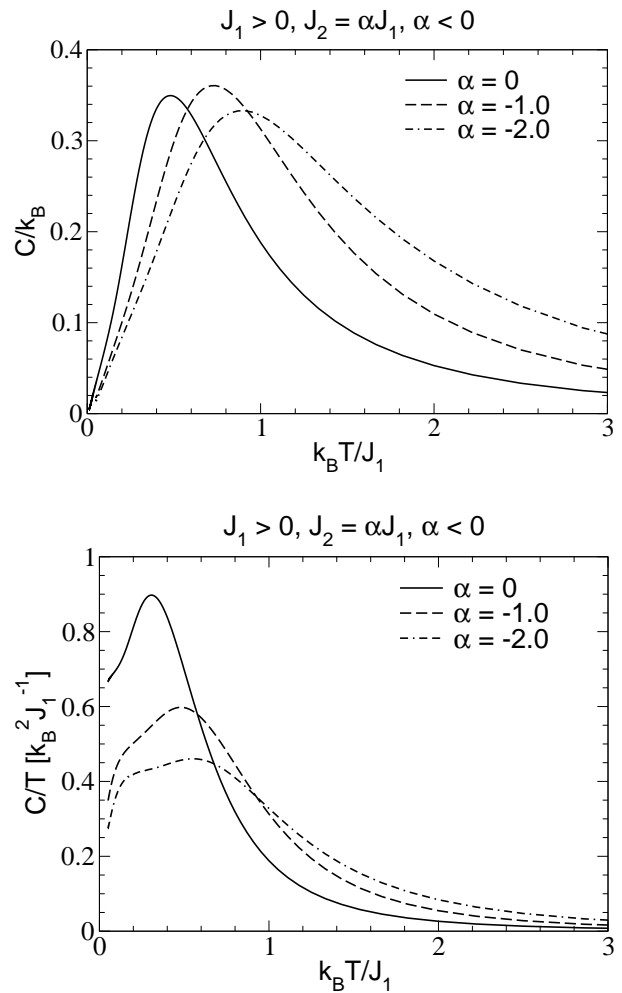

\centering
\includegraphics[width=8cm]{heat_AFF.eps}\\
\vspace{0.5cm}
\includegraphics[width=8cm]{CT_AFF.eps}
\caption{\label{C_AFF}The specific heat $C$ and heat
coefficient $C/T$ at various $\alpha$ for the AF-F case.}
\end{figure}

Figures~\ref{chi_AFF} and \ref{C_AFF} show the TMRG results on the temperature
dependence of the susceptibility $\chi$, specific heat $C$, and heat
coefficient $C/T$ at various $\alpha$ for the AF-F case. With the decrease of
$\alpha$, the behaviors of $\chi$ and $C$ do not change qualitatively.
$\chi\rightarrow \text{const}$ (nonzero) and $C\sim T$ as $T\rightarrow 0$.
The Bonner-Fisher peak in $\chi(T)$~(Ref. \cite{BonnerFisher}) moves to higher $T$
with the decrease of $\alpha$. Phase transition is not observed. The slopes of
$C$, or the intersections of $C/T$ at zero temperature (shown in
Fig.~\ref{C_AFF}) decrease with the decrease of $\alpha$. This indicates that
the density of low-excitation states becomes smaller and the spin fluctuations
are suppressed by the NNN FM interaction.

\section{Comparison with the experimental results}
\label{sec:EXP}

In the previous two sections, we have discussed thermodynamic properties of
the $J_{1}$-$J_{2}$ model for the F-AF and AF-F cases and our numerical
results cover their whole phases. In experiments, in order to understand the
physics of edge-sharing copper chains, it is important to determine both sign
and magnitude of NN and NNN interaction coefficients.  One way for this
purpose is to measure thermodynamic quantities, such as specific heat,
susceptibility and magnetization, together with numerical fitting on these
data. Such a method has been widely used in the studies on copper chains and its
efficiency has been proved, such as in SrCuO$_{2}$,~\cite{SrCuO2}
NaCu$_{2}$O$_{3}$,~\cite{NaCuO} Ca$_{2}$Y$_{2}$Cu$_{5}$O$_{15}$,~\cite{CaYCuO}
and La$_{6}$Ca$_{8}$Cu$_{24}$O$_{41}$.~\cite{LaCaCuO} 

Recently, the magnetic susceptibility and magnetization of the edge-sharing
copper oxide Rb$_{2}$Cu$_{2}$Mo$_{3}$O$_{12}$~(Ref. \cite{Solodovnikov}) have been
measured.~\cite{Hase} The most interesting finding is that no magnetic phase
transition was observed down to 2 K. Therefore, the compound is suitable for
studying the properties of the ground state of the $J_{1}$-$J_{2}$ model. It
was proposed that at first approximation, it could be described by a F-AF
$J_{1}$-$J_{2}$ model.~\cite{Hase} We have known that for the F-FA chain, the
unusually long correlation length, which exist even at high-frustrated region,
can lead to prominent finite-size effect for the usual cluster simulations.
Therefore, as a primary step, reliable numerical results on thermodynamic
quantities free of finite-size effect on the pure $J_{1}$-$J_{2}$ model may be
essential in making a quantitative comparison between experimental and
theoretical studies. In this section, we use the TMRG method to simulate the
experimental results based on model (\ref{eq:2}).

The strategy is as follows: For a fixed $\alpha$, we first use the TMRG to
calculate the temperature dependence of the susceptibility in the units of
$J_{1}$. Then, according to the position of the peak in the numerical result
and the actual value obtained in the experiment, we can determined the value
$J_{1}$ uniquely, on the condition that the calculated peak's position
coincides with the experiment. The $g$ value is taken to be $2.03$ as in
Ref.~\onlinecite{Hase}. The following is what we observed.

\begin{figure}[t]
\centering
\includegraphics[width=8cm]{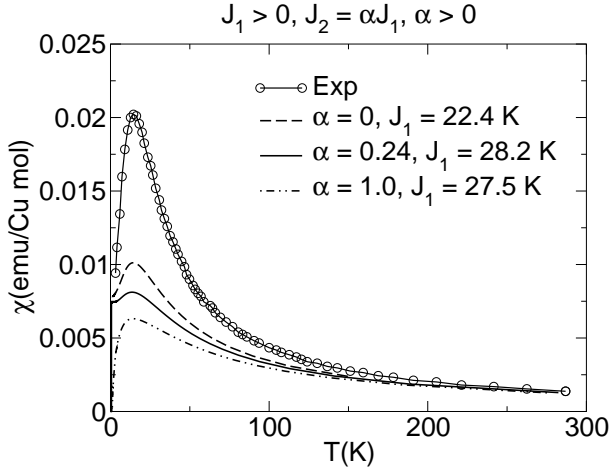}
\caption{\label{E.sus.AFAF}Comparison of the TMRG results for the
susceptibility in the AF-AF case with the experimental measurement
in Ref.~\onlinecite{Hase}.}
\vspace{0.2cm}
\end{figure}

For the AF-AF case (Fig.~\ref{E.sus.AFAF}), the height of the peak
decreases with the increase of $\alpha$. The possibility for this
type of interaction can be excluded.

\begin{figure}[t]
\centering
\includegraphics[width=8cm]{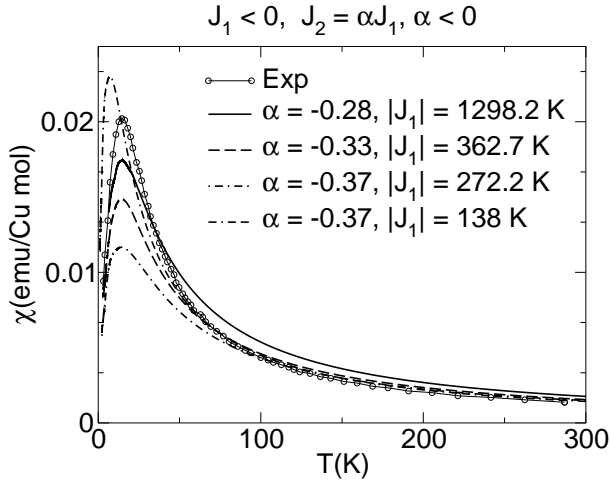}
\caption{\label{E.sus.FAF}Comparison of the TMRG results for the
susceptibility in the F-AF case with the experimental measurement
in Ref.~\onlinecite{Hase}.} \vspace{0.2cm}
\end{figure}

For the F-AF case (Fig.~\ref{E.sus.FAF}), we see that with $\alpha$
approaching the critical point $-1/4$, to keep the peak position unchanged,
the magnitude of $J_{1}$ acquires a value of thousands of Kelvin. Furthermore,
the deviation from the experiment data at high temperatures becomes more
conspicuous. While taking the parameters $J_{1}=-138$ K and $\alpha=-0.37$ as
in Ref.~\onlinecite{Hase}, the temperature for the peak is less than
$T_{\text{max}}=14.3$ K found in the experiment. Besides, for $\alpha<-0.5$, a
broad peak located at $T\sim J_{2}$ in the temperature dependence of
susceptibility emerges, which is a characteristic feature of one-dimensional
Heisenberg antiferromagnetic chain.\cite{BonnerFisher} This suggests that the
antiferromagnetic component becomes significant for larger $\alpha$.  However,
we failed to obtain a satisfactorily stable result below the peak temperature
when $\alpha<-1$.

\begin{figure}[t]
\centering
\includegraphics[width=8cm]{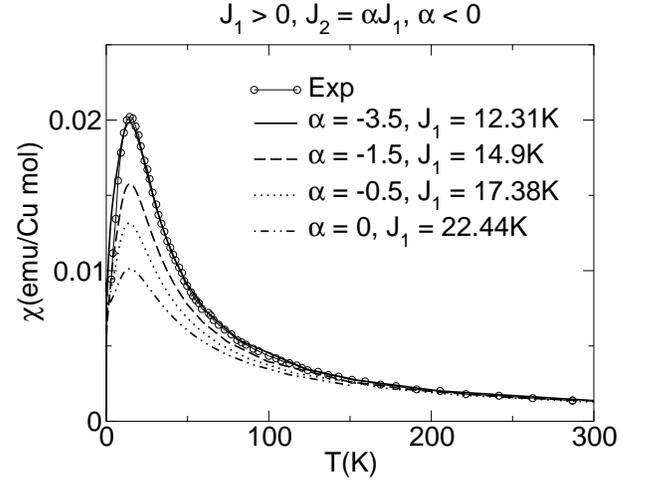}
\caption{\label{E.sus.AFF}Comparison of the TMRG results for the
susceptibility in the AF-F case with the experimental measurement
in Ref.~\onlinecite{Hase}.} \vspace{0.4cm}
\end{figure}

\begin{figure}[t]
\centering
\includegraphics[width=8cm]{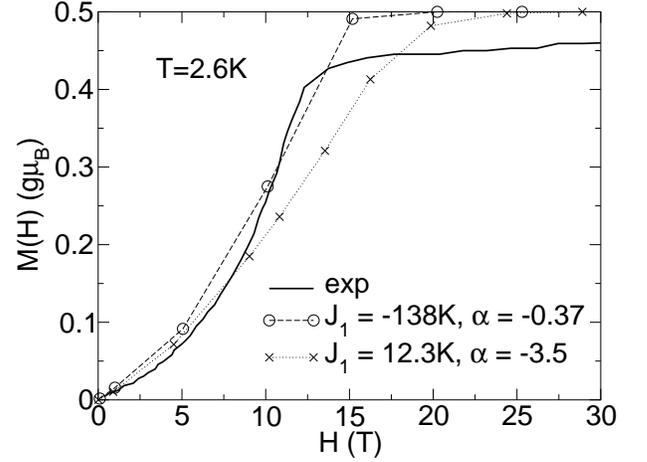}
\caption{\label{Mag}Comparison of the TMRG results for the
magnetization in the F-AF (circles) and AF-F (crosses)
cases with the experimental measurement
in Ref.~\onlinecite{Hase}.}
\vspace{0.2cm}
\end{figure}

For the AF-F case (Fig.~\ref{E.sus.AFF}), at $J_{1}=12.31$ K and
$\alpha=-3.5$, the numerical result seems to fit the experiment data rather
well, except for a little deviation in the left side of the peak at low
temperatures. But there are several questionable points remained. According to
the analysis of Mizuno {\emph et al}~\cite{Mizuno}, the NNN interaction through
Cu-O-O-Cu should be antiferromagnetic, i.e.  $J_{2}>0$. However, the
interactions of AF-F type may also be a candidate. In
Ref.~\onlinecite{LaCaCuO3}, Matsuda and coworkers proposed a model to explain
the anomalous magnetic excitations in the edge-sharing CuO$_{2}$ chains of
La$_{5}$Ca$_{9}$Cu$_{24}$O$_{41}$. The intra- and interchain interactions are
of AF-F type. (See also Fig.~\ref{Mag}.)

We also calculated the field dependence of the magnetization for the two
cases: $J_{1}=-138$ K, $\alpha=-0.37$; and $J_{1}=12.3$ K, $\alpha=-3.5$, to
verify further if the only $J_{1}$-$J_{2}$ model is sufficient to describe the
behaviors of the compound. We find the anomalous slow saturation of the
magnetization cannot be reproduced either, and the result for the AF-F case
seems worse.

From the TMRG numerical results on the $J_{1}$-$J_{2}$ model and the
comparison made with the experiment, we feel that only the NN and NNN
interactions can not describe the properties of
Rb$_{2}$Cu$_{2}$Mo$_{3}$O$_{12}$ satisfactorily, although a model based on the
F-AF interactions seems to be an appropriate starting point.

The above facts reveal the complexity of the interactions underlying the
edge-sharing copper oxides. On one hand, due to the strong electron
correlations in these so-called Mott insulators, very limited information on
the electronic structure can be obtained reliably. This makes it difficult to
calculate accurately the superexchange interaction For instance, based on a
three-band Hubbard Hamiltonian and cluster calculation, the NN and NNN
interaction $J_{1}$, $J_{2}$ for Li$_{2}$CuO$_{2}$ were obtained as
$J_{1}=-100$ K, $J_{2}=62$ K ($\alpha=-0.62$).\cite{Mizuno} The results of the
quantum chemical calculation were given as $J_{1}=-142$ K, $J_{2}=22$ K
($\alpha=-0.15$).\cite{Li2CuO2_3} On the other hand, because the NN coupling
in edge-sharing copper oxides is extremely small, other interactions, such as
quantum frustrations, weak interchain correlations, and anisotropies can all
have a chance to play an unnegligible role in determining the phase and
behavior of the system. They are also closely related to the lattice
structures and chemical compositions. These combined effects make any
reasonable analysis intricate. In order to quantitatively recover the
experimental data and various magnetic orders at low temperatures, more
parameters are needed. This brings some ``flexibility'' to the theory. For
instance, the broadening of the magnetic excitations found in
Ca$_{2}$Y$_{2}$Cu$_{5}$O$_{10}$ requires the introduction of the
antiferromagnetic interchain interactions and anisotropies for superexchange
interactions.\cite{CaYCuO2} In order to understand the helicoidal magnetic
order in NaCu$_{2}$O$_{2}$, four parameters including frustrated longer-range
exchange interactions are needed.\cite{NaCuO} However, we wish to emphasize
here that accurate knowledge on the behavior of model~\ref{eq:2}, especially
at low-temperature region, is indispensable in understanding the properties of
these materials.

In conclusion, we go back to the compound
Rb$_{2}$Cu$_{2}$Mo$_{3}$O$_{12}$ and take a closer look at the
lattice structure of Rb$_{2}$Cu$_{2}$Mo$_{3}$O$_{12}$. Since the
nearest neighbor Cu-Cu bond has two slightly alternating
configurations by turns, the chain is distorted into a zigzag shape
(See Fig.~1 in Ref.~\onlinecite{Hase}). Additional antisymmetric
exchange interactions, such as Dzyaloshinskii-Moriya (DM)\cite{DM}
or anisotropic interactions,~\cite{Kataev01} together  with an
alternating $g$ tensor, should become more important than in a
straight-line chain. As discussed by Dzyaloshinskii and
Moriya\cite{DM} for the magnetic crystals with lower symmetries, the
effect of this antisymmetric exchange should become more manifest
than the exchange anisotropy.  The magnitude of this interaction is
estimated as
\begin{equation}
D\sim\left(\Delta g/g\right)J \,,
\end{equation}
and the usual term of DM interaction can be expressed as
\begin{equation}
H_{\text{DM}} = \sum_{j}\mathbf{D}_{j}\cdot
\left(\mathbf{S}_{j}\times\mathbf{S}_{j+1}\right) \,.
\end{equation}

Because of the alternating $g$ tensor and the DM interaction, an
external uniform magnetic field can induce an effective staggered
field. If $\mathbf{D}_j$ takes the form $(-1)^{j}\mathbf{D}$, as it
should be in the present case, the transverse component of the
staggered field, which is perpendicular to the uniformly applied
field, becomes dominant. For the $S=1/2$ Heisenberg
antiferromagnetic chain, a gap is generated by a staggered
field,\cite{Oshikawa} and the magnetization becomes gradually
saturated for large fields.\cite{Jize} The similar mechanism may
also work for the F-AF case. A combination with the frustration
effect caused by the NNN $J_2$ make the situation more interesting.

Recently, it was suggested that for $\alpha<-0.38$, there exists an
incommensurate-commensurate transition at some critical field in the
magnetization process.\cite{Dmitriev} The sharp increase up to $M\simeq0.4$ at
$B\simeq14$ T and the following gradual saturation found in the experiment is
argued to be connected to this phase transition.

\section{Summary}
\label{sec:SUM}

In this paper, we explored the properties of the zigzag spin chain
with different combinations of ferro- and antiferromagnetic
interactions between the NN and NNN sites. The existence of the gap
in the F-AF case and the nonexistence of the gap in the AF-F case
were discussed. Thermodynamic properties of the zigzag spin chain in
various phases were studied by using the TMRG method. The obtained
results were used to compare with the experimental data from
Rb$_{2}$Cu$_{2}$Mo$_{3}$O$_{12}$. We pointed out that for such
edge-sharing copper oxide chains, besides the NN and NNN couplings,
more ingredients, such as the interchain or DM exchange
interactions, may be important in these materials.

\begin{acknowledgments}
This work was supported by the National Natural Science Foundation of China.
\end{acknowledgments}

\end{document}